\documentclass{cimento}

\newcommand{\met}{\ensuremath{{\slash\kern-.7emE}_{T}}}
\newcommand{\mt}{\ensuremath{m_T}}
\newcommand{\pte}{\ensuremath{p_T^e}}

\usepackage{graphicx}  % got figures? uncomment this
\title{$W$ mass and width measurements at the Tevatron}
\author{Alex Melnitchouk for the CDF and D0 Collaborations\from{ins:x}}
\instlist{\inst{ins:x}University of Mississippi, University, Mississippi, 38677, USA}
\PACSes{
\PACSit{14.70.Fm}{W bosons}
\PACSit{13.38.Be}{Decays of W bosons}}
\begin{document}
 
\maketitle

\begin{abstract}
Most recent results of $W$ boson mass and width measurements performed by 
CDF and D are reported. 
at the center-of-mass energy of 1.96 TeV. 
Integrated luminosity ranges from 0.2 fb$^{-1}$ to 1.0 fb$^{-1}$ depending on the analysis.
\end{abstract}

\section{Introduction}
Measurement of the $W$ boson mass ($M_{W}$) 
provides us with a uniquely powerful key to uncovering the origin of the electroweak
symmetry breaking and learing about new physics. 
A precision measurement of $M_{W}$ is one of the highest priorities for the Tevatron experiments.
$M_{W}$ measurement combined with precise measurement of the top quark mass ({$M_{top}$), constrains the mass of the Higgs boson.

On the other hand, the width of the 
$W$ boson ($\Gamma_{W}$) is expected to be insensitive to new physics. 
Therefore its precise measurement 
is very important for improving the experimental knowledge of the Standard Model.
Currently CDF\cite{cdf_det} and D0\cite{d0_det} provide most precise derect
measurements of both $M_{W}$\cite{d0mw,cdfmw} and $\Gamma_{W}$\cite{d0gw,cdfgw}
For these measurements CDF uses both electron and muon decay channels of the $W$,
while D0 uses only electron channel.

\section{Identification of Electrons and Muons}
Electrons are identified as an electromagnetic (EM) cluster 
reconstructed with a simple cone algorithm. 
To reduce the background of jets faking electrons, 
electron candidates are required to have a large fraction 
of their energy deposited in the EM section of the calorimeter 
and pass energy isolation and shower shape requirements.
Electron candidates are classified as \textit{tight} 
if a track is matched spatially to EM cluster and if the track 
transverse momentum is close to the transverse energy of the EM cluster.
In CDF electrons are reconstructed both in the central
calorimeter and plug calorimeter ($|\eta| < 2.8$) 
while electrons in D0 are reconstructed 
in the central and endcap calorimeters ($|\eta|<1.05$ and $1.5<|\eta|<3.2$). 
Here $\eta = -\ln\tan(\theta/2$, and  $\theta$ is the polar angle with respect 
to the proton direction. Both CDF and D0 require  \textit{tight} electrons 
in the central calorimeter ($|\eta|<1.05$) for $W \rightarrow e\nu$ 
candidates. Electron energies are measured with the calorimeter,
while electron direction is measured with tracking detectors, using
tracks that are matched to electron cluster in the calorimeter.

Muons are identified by a track in the muon system matched to a track 
in the central tracking system.
Measurements include the muons reconstructed in the central muon extension sub-detector which 
extends the coverage from $|\eta|<0.6$ to $|\eta|<1.$

\section{W Mass}
$M_{W}$ is measured using three transverse kinematic variables: 
the transverse mass $m_{T} = \sqrt {2p_{T}^{e,\mu}p_{T}^{\nu}(1 - \cos\Delta\phi)}$, 
the lepton ( $p_{T}^{e,\mu}$) and neutrino ($p_{T}^{\nu}$) 
transverse momentum distributions, where $\Delta\phi$ is the opening angle 
between the electron(muon) and neutrino momenta in the plane transverse to the beam. 
Neutrino transverse momentum ($p_{T}^{\nu}$) is inferred from the imbalance
of transverse energy. We also call it missing $E_{T}$ (MET).

A sophisticated parametrized fast Monte Carlo simulation is used 
for modeling these variables as a function of $M_{W}$.
Fast simulation includes models of electron, recoil system, and backgrounds.
Electron efficiencies, resolution and energy scale parameterizations are tuned
to $Z \rightarrow ee$ data.
Recoil system represents energy deposited in the calorimeter from all
sources except the electron(s). Recoil system consists of three major components:
hard recoil (particles that collectively balance the $p_{T}$ of the W of Z boson),
underlying event, and additional interactions. Contribution from the third component depends 
on the instantaneous luminosity. Hard recoil is modeled using full detector simulation,
while the other two componenets are described by real data events.
Full recoil model is tuned to $Z \rightarrow ee$ data, using imbalance between
Z boson momentum measured with electrons and with recoil system.
Sources of backgrounds to $W \rightarrow e\nu$ events include $W \rightarrow \tau\nu \rightarrow e\nu\nu$,
QCD, and $Z \rightarrow ee$ prccesses.

$M_{W}$ is extracted from a binned maximum-likelihood fit between the data and simulation.
$\Gamma_{W}$ is measured with $m_{T}$ variable using the same analysis framework
as $M_{W}$. Fig.~\ref{fig:mwfig} shows a comparison between data
and fast simulation. It also shows final $M_W$ results from D0 and CDF along with
other $M_W$ measurements and combinations
D0 result agrees with the world average and the individual measurements and
is more precise than any other $M_W$ measurement from a single measurement.
Fig.~\ref{fig:cdfmwfig} shows comparison between data and fast simulation for CDF $M_W$ measurement.
\begin{figure}[hbpt]
\begin{center}
  \includegraphics[width=.495\textwidth]{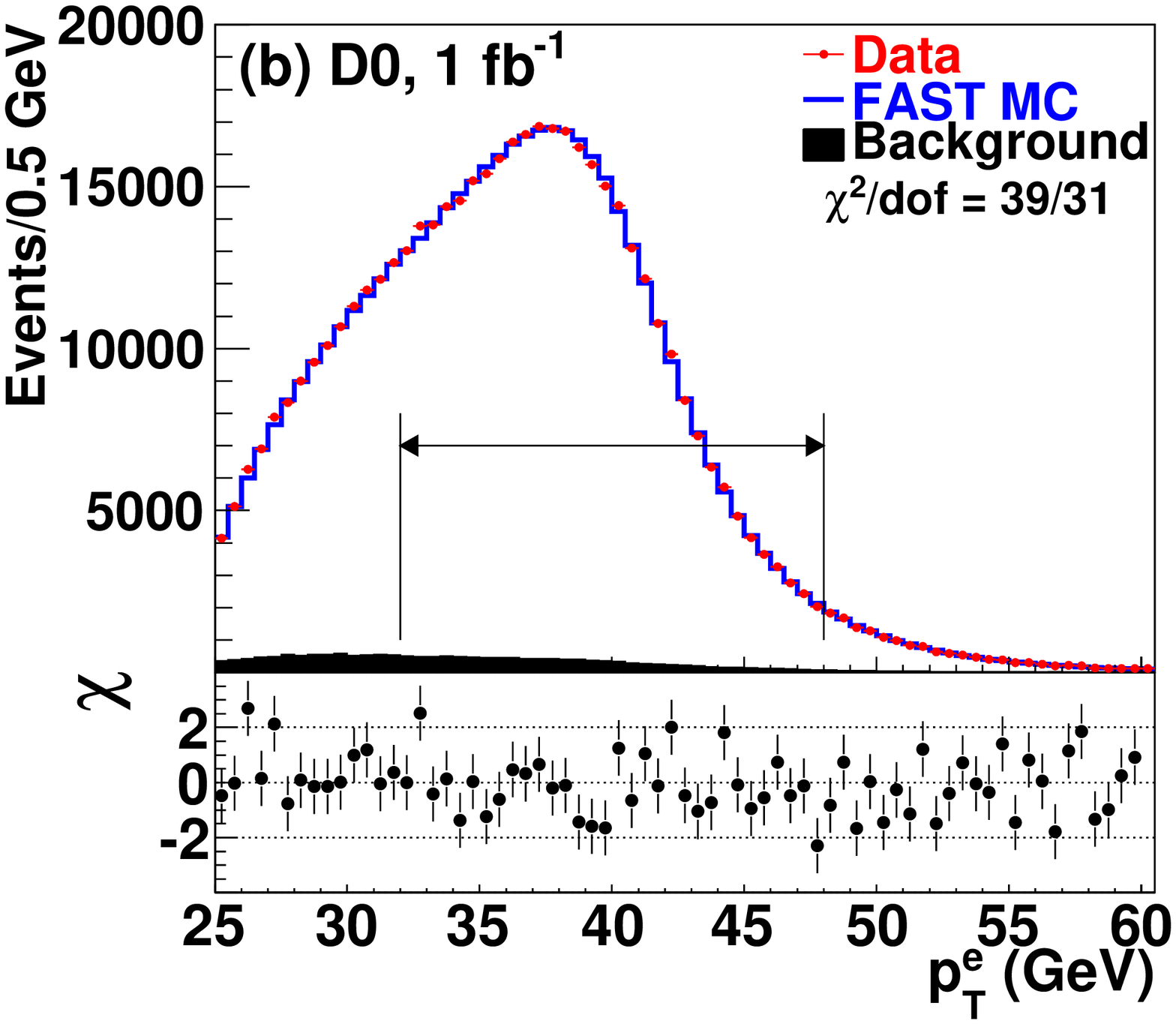}
  \includegraphics[width=.495\textwidth]{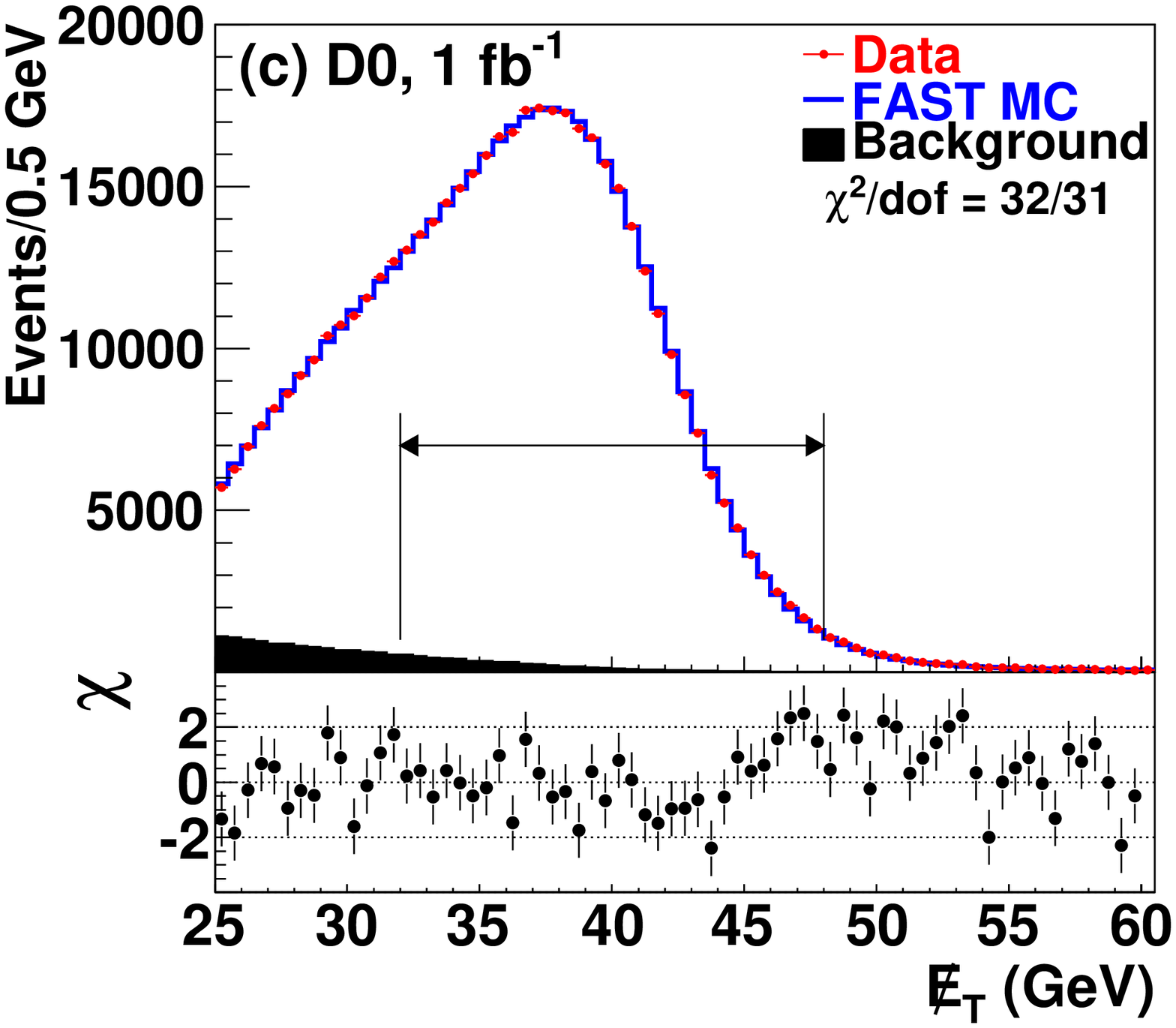}
  \includegraphics[width=.495\textwidth]{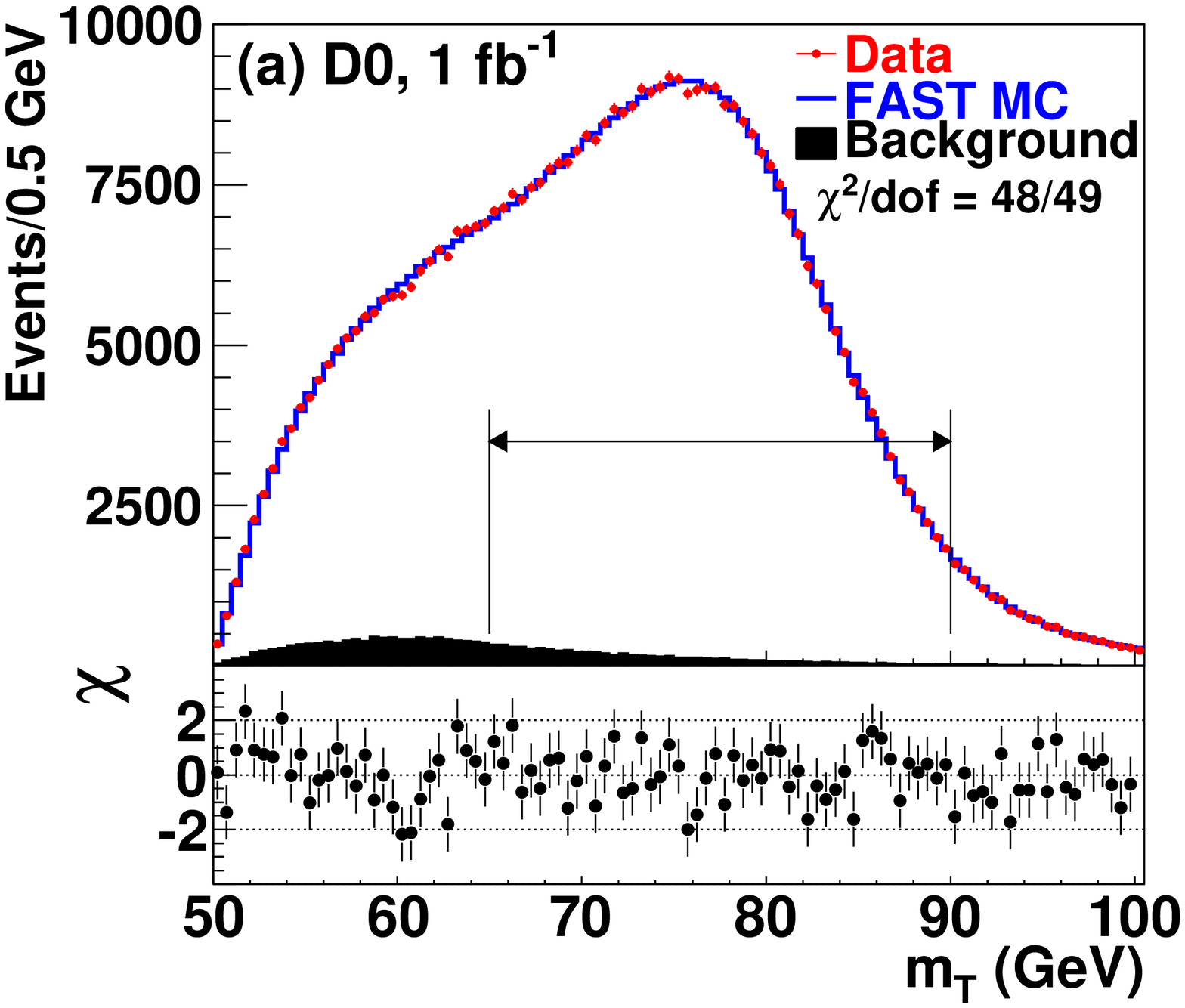}
  \includegraphics[width=.495\textwidth]{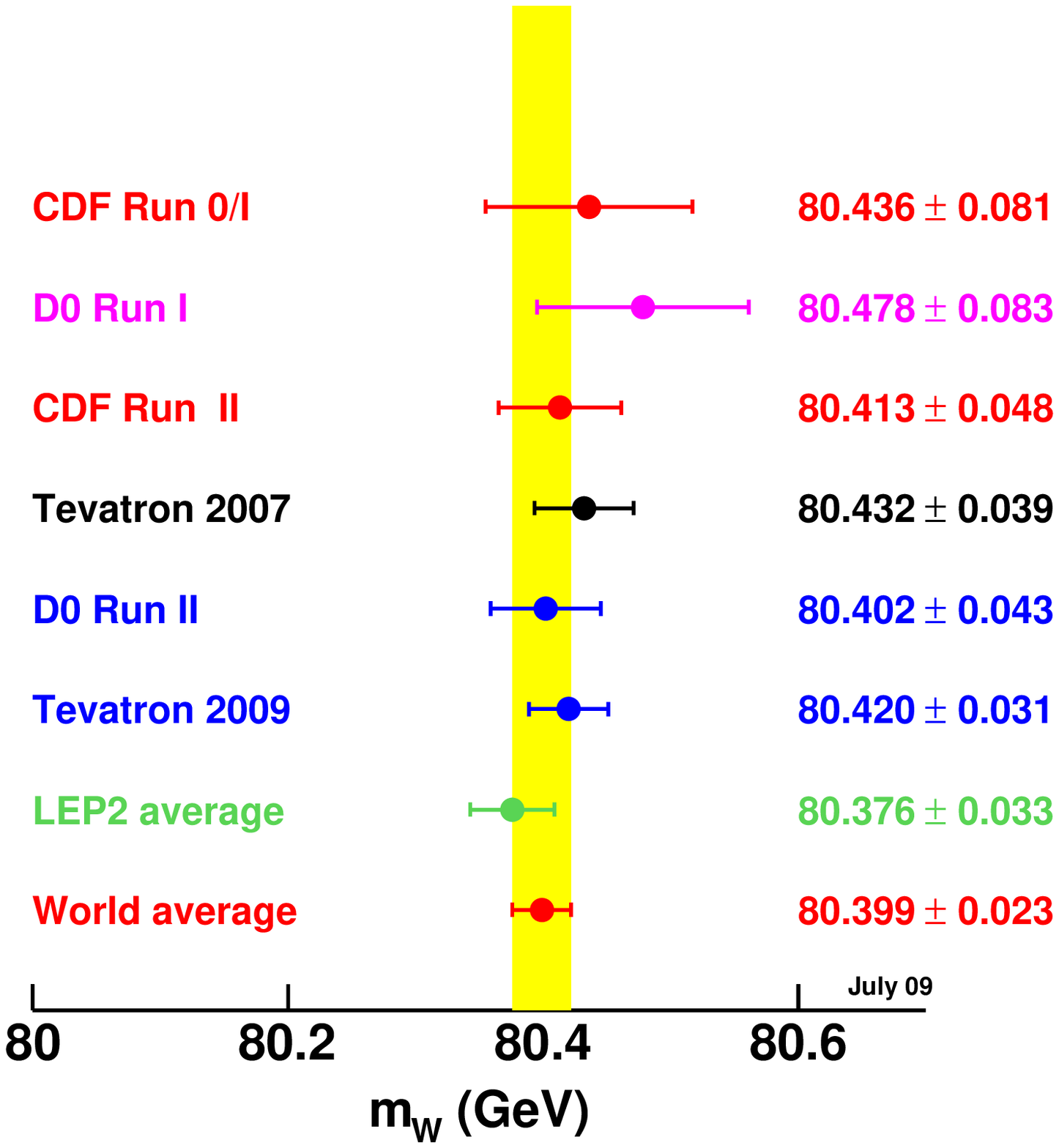}
 \end{center}
\vspace{-1pc}
  \caption{
Top left, top right, and bottom left show
electron $p_T$, $m_{T}$, and MET distributions in $W \rightarrow e\nu$ data and fast simulation ({\sc fastmc}).
Added background is shown as well. Signed $\chi$ distributions are shown in the bottom of part of each plot.
Signed $\chi$ is defined as
$\chi_i = [N_i-\,(${\sc fastmc}$_i)]/\sigma_i$ 
for each point in the distribution, $N_i$ is
  the data yield in bin $i$ and $\sigma_i$ is the 
  statistical uncertainty in bin $i$.\newline
Bottom right: summary of the measurements of the $W$ boson mass and their average. 
The result from the Tevatron corresponds to the values which includes 
corrections to the same W boson width and PDFs. 
The LEP II results are from \cite{lep}. 
An estimate of the world average of the Tevatron and LEP 
results is made assuming no correlations between the Tevatron and LEP 
uncertainties.
}
\label{fig:mwfig}
\end{figure}
\begin{figure}[hbpt]
\begin{center}
\includegraphics[width=.495\textwidth]{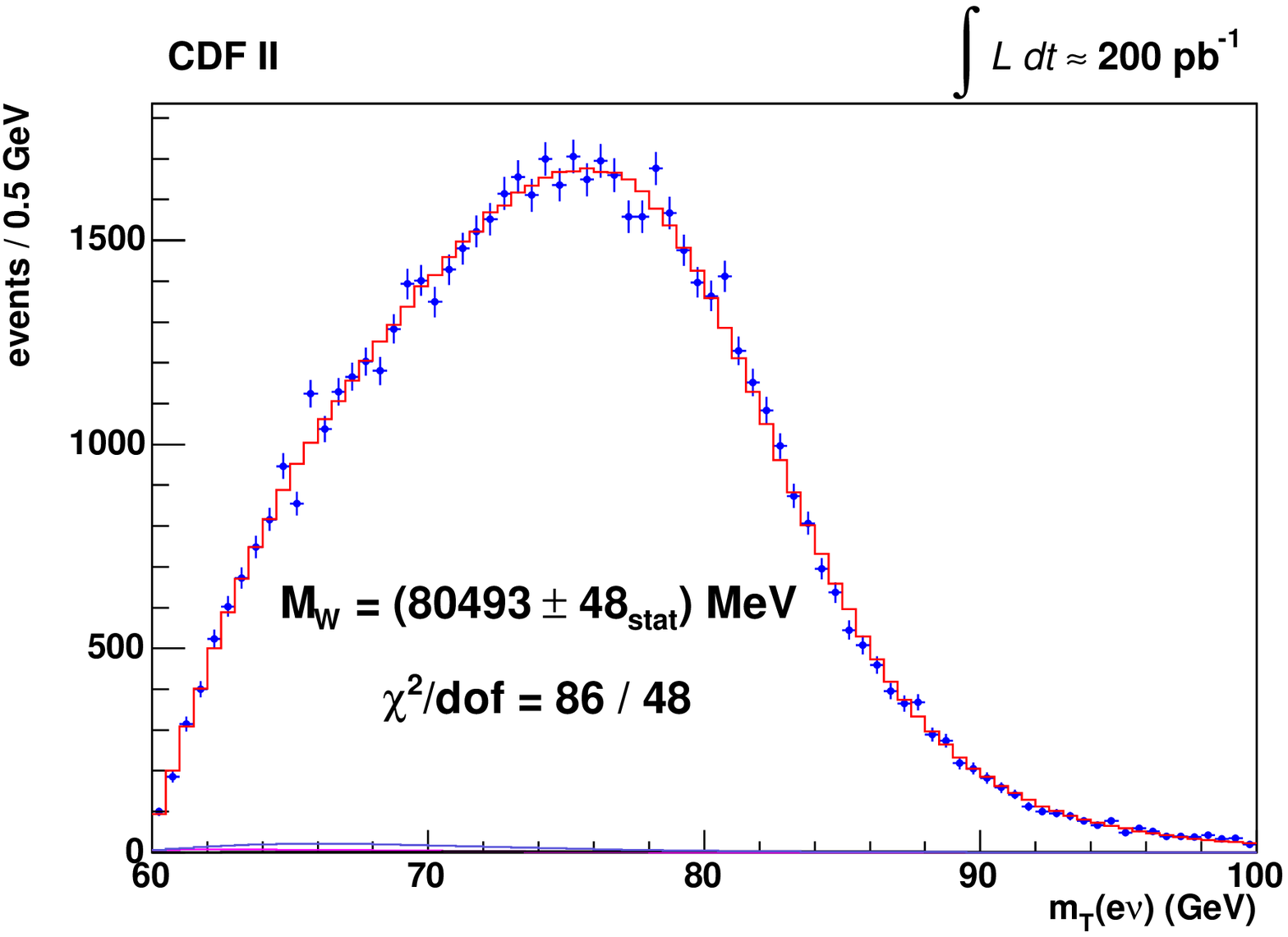}
\includegraphics[width=.495\textwidth]{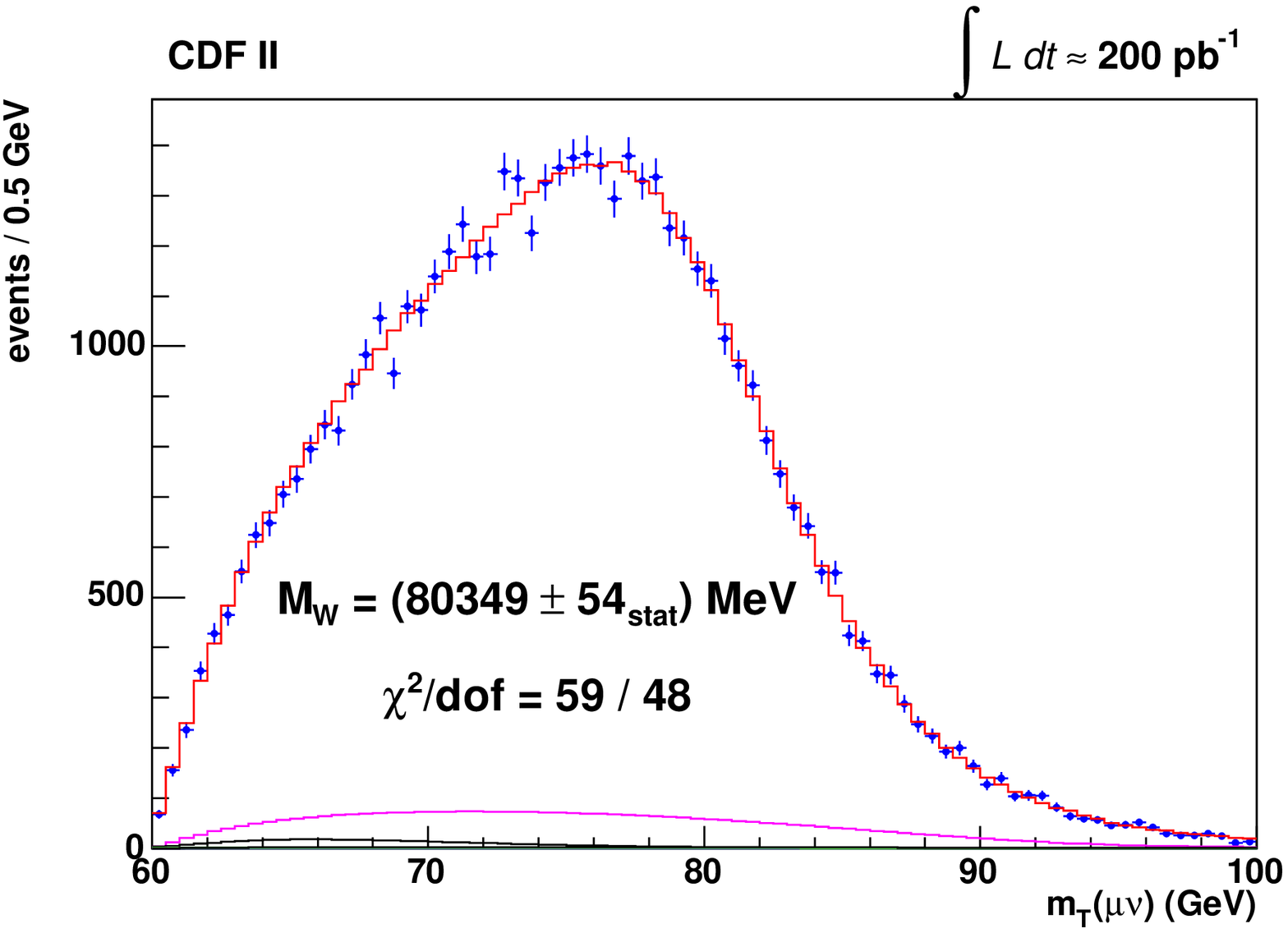}
\includegraphics[width=.495\textwidth]{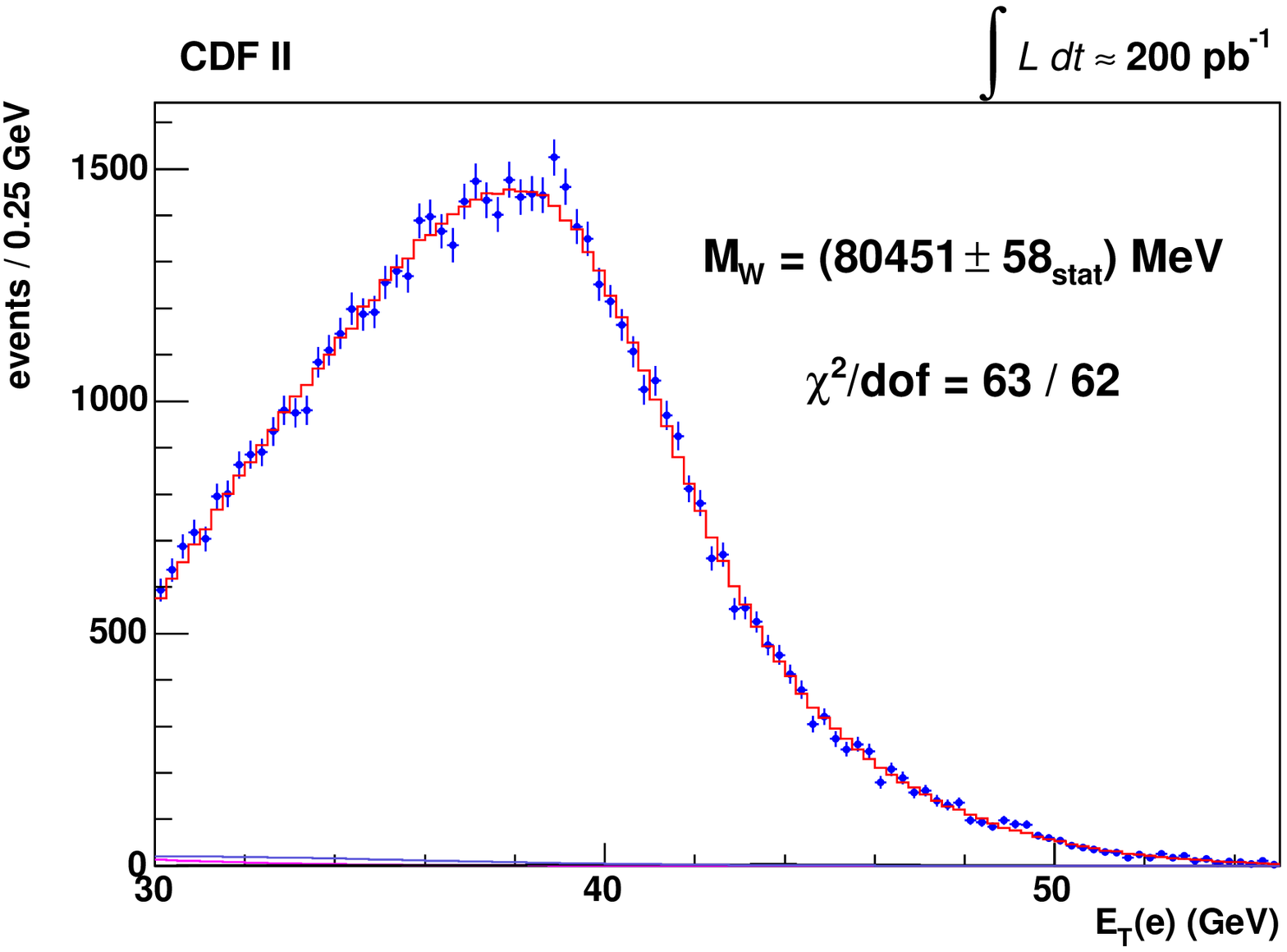}
\includegraphics[width=.495\textwidth]{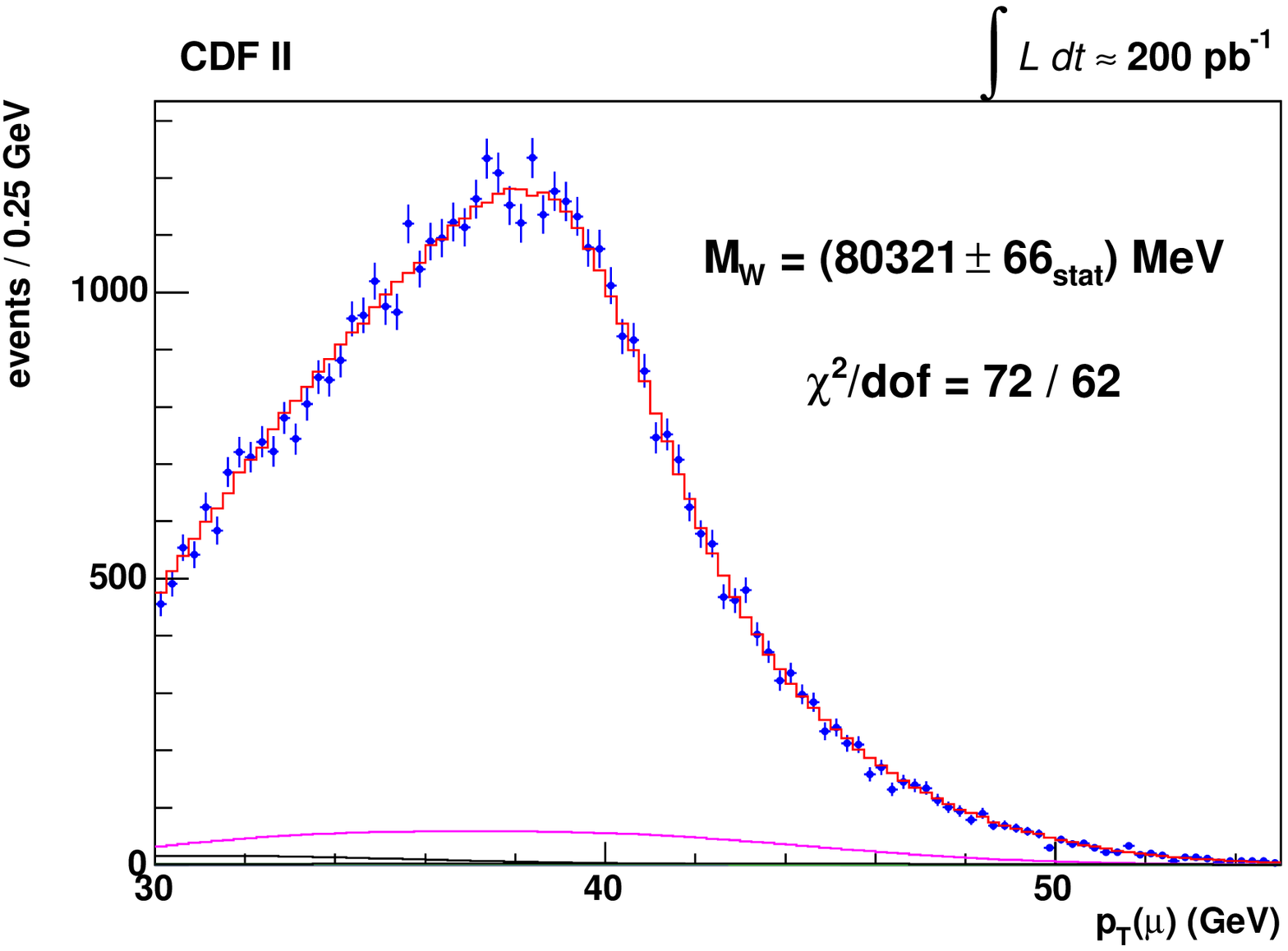}
\includegraphics[width=.495\textwidth]{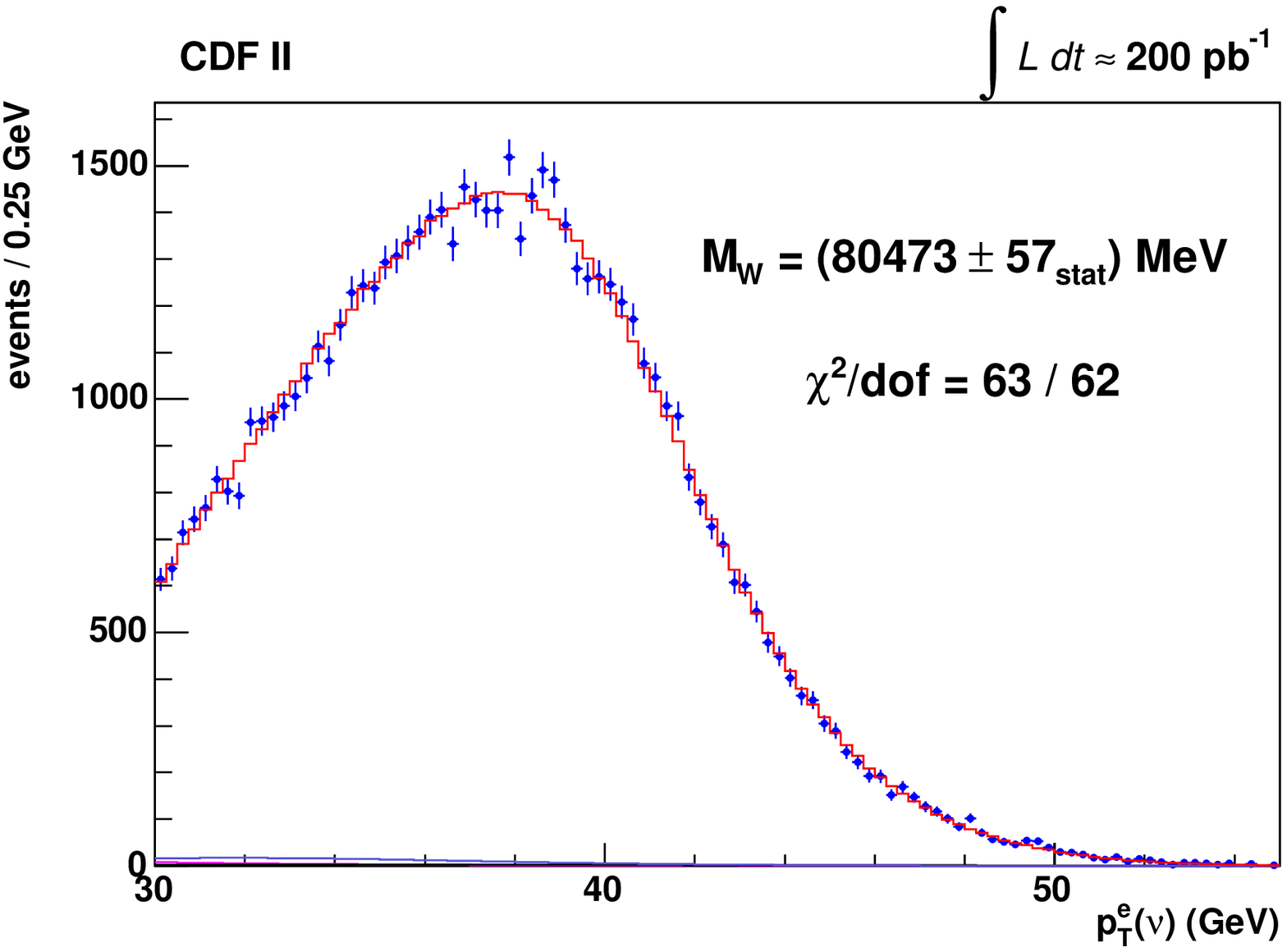}
\includegraphics[width=.495\textwidth]{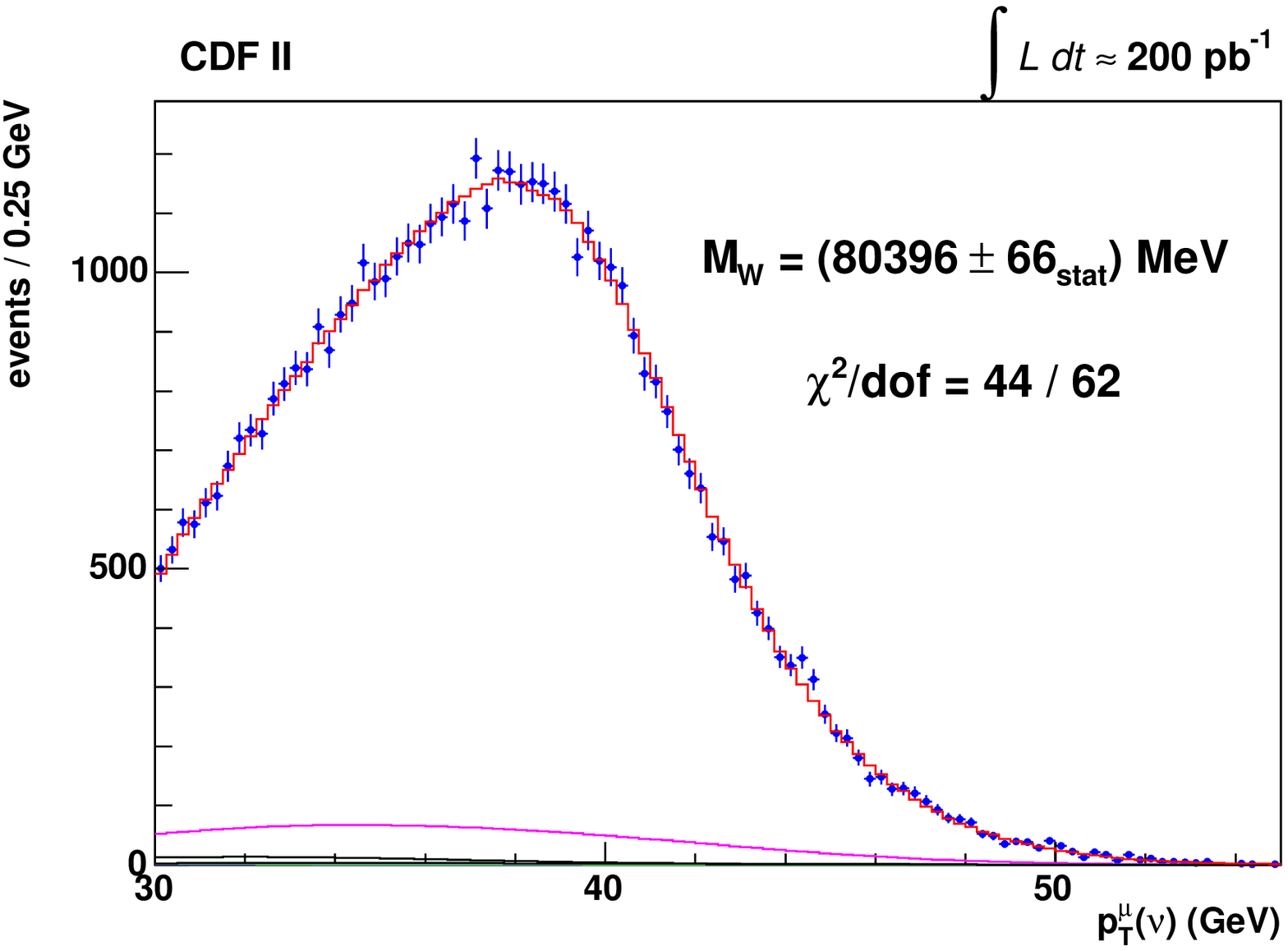}
\end{center}
\vspace{-1pc}
  \caption{
Distibutions of $M_W$ observables in CDF measurement.
Blue -- data. Red -- fast simulation. Fit results and
statistical errors are indicated.
Left column: electron channel. Right column: muon channel.
Top row: $m_{T}$. Middle row: charged lepton $p_{T}$. 
Bottom row: neutrino $p_{T}$.}
\label{fig:cdfmwfig}
\end{figure}

Dominant uncertainties in $M_W$ measurements come 
from lepton energy scale measurements. To first order
fractional error on the lepton energy scale translates to fractional error 
on the W mass\cite{ashutoshandjan}. 

D0 determines electron energy scale
using high $p_{T}$ electrons from $Z \rightarrow ee$ decays.
Precision of such calibration is limited mostly by the size of
the $Z \rightarrow ee$ sample.

CDF relies on tracking detector for both
electron and muon energy scale calibration. First tracking detector
is calibrated using $J/\psi \rightarrow \mu\mu$ events. $J/\psi$
invariant mass is measured as a function of muon momentum.
Fig.~\ref{fig:cdfjpsimass} shows the correction needed
to make measured $J/\psi$ mass to be at its PDG value (overall offset) and 
independent of muon momentum (slope). This correction was implemented
in the simulation by adjusting the energy-loss model. Then tracker calibration
is transported to the calorimeter using $W \rightarrow e\nu$ electrons
near the peak of the E/p distribution, shown also in Fig.~\ref{fig:cdfjpsimass}.
Tables \ref{tab:mwtable} and \ref{tab:cdfmwtable} show uncertainties for
$M_W$ measurements by D0 and CDF respectively.
\begin{table}
\vspace{0.55cm}
  \caption{Uncertainties of D0 $M_W$ Measurement (MeV).}
  \label{tab:mwtable}
  \begin{tabular}{rccl}
    \hline
   Source                          &$\mt$ & $\pte$ &  $\met$\\
  \hline 
  Experimental             &    &     & \\ \hline
  Electron energy calibration       & 34 &  34 & 34 \\
  Electron resolution model         &  2 &   2 &  3 \\
  Electron energy offset            &  4 &   6 &  7 \\
  Electron energy loss model        &  4 &   4 &  4 \\
  Recoil model                      &  6 &  12 & 20 \\
  Electron efficiencies             &  5 &   6 &  5 \\
  Backgrounds                       &  2 &   5 &  4 \\ \hline
  Experimental Subtotal             & 35 &  37 & 41 \\ \hline
  Production Model                  &    &     & \\ \hline
  PDF                               &  10 &  11 & 11 \\
  QED                               &  7 &   7 &  9 \\
  Boson $p_T$                       &  2 &   5 &  2 \\ \hline
  Production Model Subtotal         & 12 &  14 & 14 \\ \hline
  Statistical                       &  23 & 27 & 23 \\ \hline
  Total                             &  37 & 40 & 43 \\ 
  \end{tabular}
\end{table}
\begin{table}
\vspace{-0.45cm}
  \caption{Uncertainties of CDF $M_W$ Measurement (MeV).}
  \label{tab:cdfmwtable}
  \begin{tabular}{rccl}
    \hline
   Source                          &$\mt$ & $\pte$ &  $\met$\\
   \hline
   $$                          & $e$,$\mu$,common & $e$,$\mu$,common & $e$,$\mu$,common\\
  \hline
  Lepton Scale                &  30,17,17 &  30,17,17 &  30,17,17 \\
  Lepton Resolution           &  9,3,0    &  9,3,0    &  9,5,0   \\
  Recoil Scale                &  9,9,9    &  17,17,17 &  15,15,15 \\
  Recoil Resolution           &  7,7,7    &  3,3,3    &  30,30,30    \\
  $\mbox{U}_{||}$ Efficiency  &  3,1,0    &  5,6,0    &  16,30,0 \\
  Lepton Removal              &  8,5,5    &  0,0,0    &  16,10,10 \\
  Backgrounds                 &  8,9,0    &  9,19,0   &  7,11,0 \\
  $p_{T}(W)$                  &  3,3,3    &  9,9,9    &  5,5,5 \\ 
  PDF                         &  11,11,11 &  20,20,20 &  13,13,13 \\
  QED                         &  11,12,11 &  13,13,13 &  9,10,9 \\ \hline
  Total Systematic            &  39,27,26 &  45,40,35 &  54,46,42 \\ \hline
  Statistical                 &  48,54,0  &  56,68,0  &  57,66,0 \\  \hline
  Total                       &  62,60,26 &  73,77,35 &  79,80,42 \\ 
  \end{tabular}
\end{table}
\begin{figure}[hbpt]
\begin{center}
\includegraphics[width=.495\textwidth]{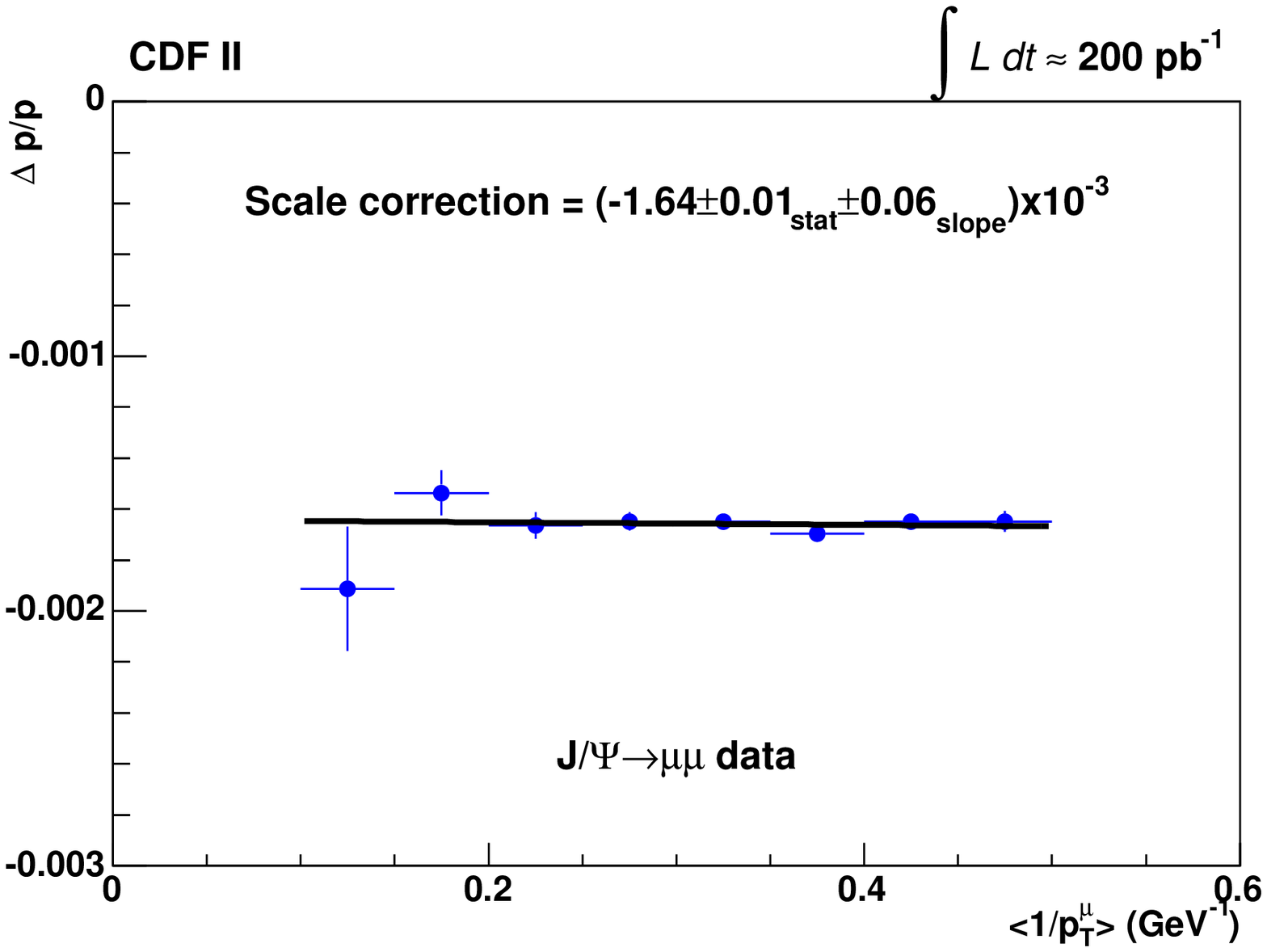}
\includegraphics[width=.495\textwidth]{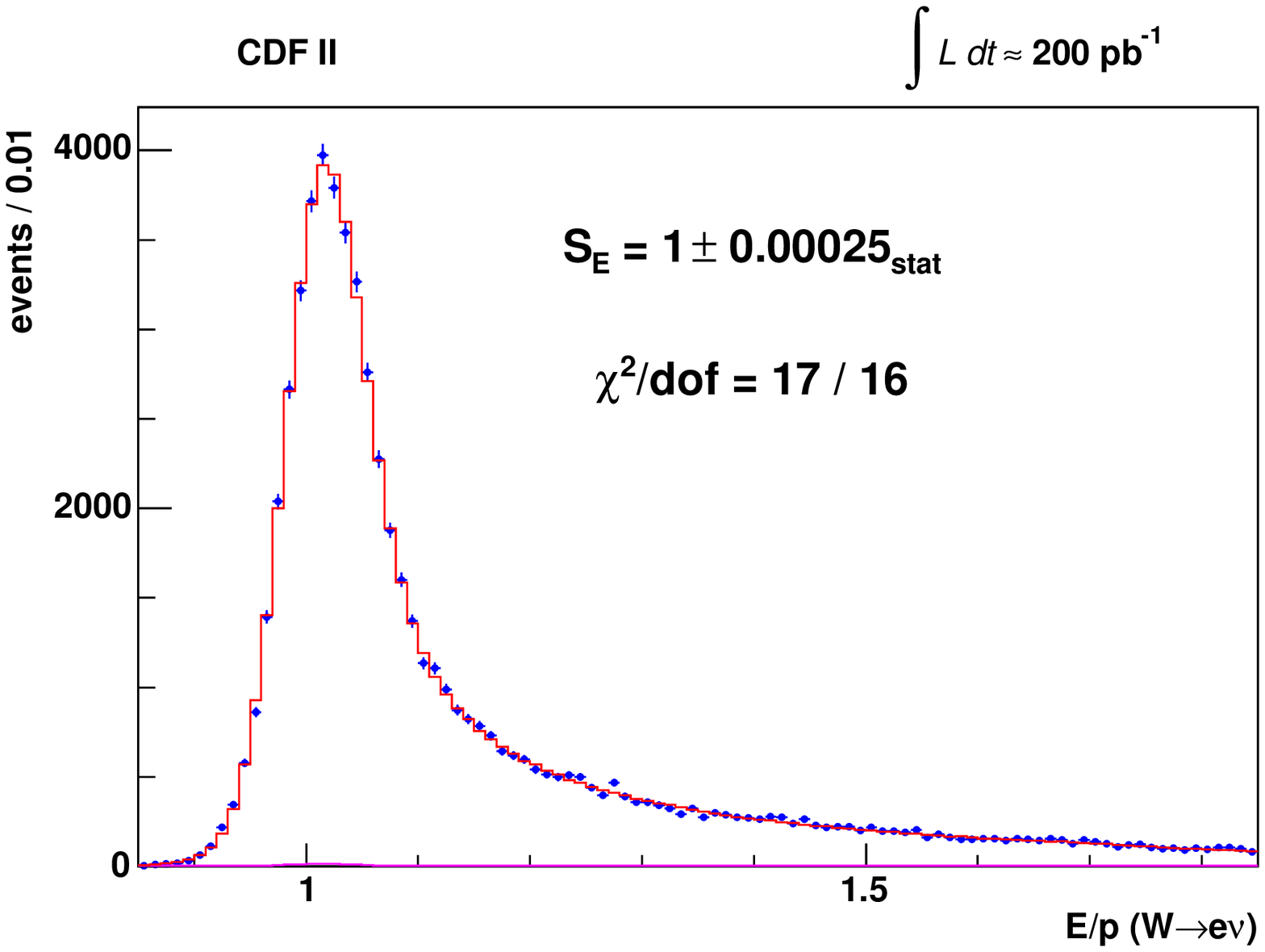}
\end{center}
\vspace{-1pc}
 \caption{
Left: fractional muon momentum correction as a function of inverse momentum.
Right: ratio of electron energy measured in the calorimeter
to electron momentum measured by the tracking system in $W \rightarrow e\nu$ events.}
\label{fig:cdfjpsimass}
\end{figure}

\section{W Width}
Although $M_W$ and $\Gamma_{W}$ measurements are performed with the same method
and both rely on $m_{T}$ distribution, they are mostly sensitive to
different features of the latter. $M_W$ is mostly sensitive to the
position of the Jacobian peak. $\Gamma_{W}$ is mostly sensitive to the tail
of the $m_{T}$ distribution. To first order $\Gamma_{W}$ is proportional to the fraction
of events in the tail.  Fit for $\Gamma_{W}$ is performed in 
the high $m_{T}$ tail region (90-200 GeV for both CDF and D0).
This region is sensitive to the Breit-Wigner line-shape 
and less sensitive to the detector resolution.

Fig.~\ref{fig:d0andcdfwidth} shows $m_T$ distributions
from CDF and D0 as well as final results compared with other measurements and 
combinations. 
D0 result is $\Gamma_{W} = 2.028 \pm 0.039 \textrm{(stat)} \pm 0.061 \textrm{(syst)} = 2.028 \pm 0.072$ GeV. 
CDF result is $\Gamma_{W} = 2.032 \pm 0.045 \textrm{(stat)} \pm 0.057 \textrm{(syst)} = 2.032 \pm 0.073$ GeV. 
Combined Tevatron average is $\Gamma_{W} = 2.046 \pm 0.049$ GeV \cite{gwcombpaper}. 
Tables \ref{tab:wgtable} and \ref{tab:cdfwgtable} give the detailed breakdown of 
uncertainties for $\Gamma_W$ measurements at D0 and CDF.  
\begin{figure}[hbpt]
\begin{center}
\includegraphics[width=.495\textwidth]{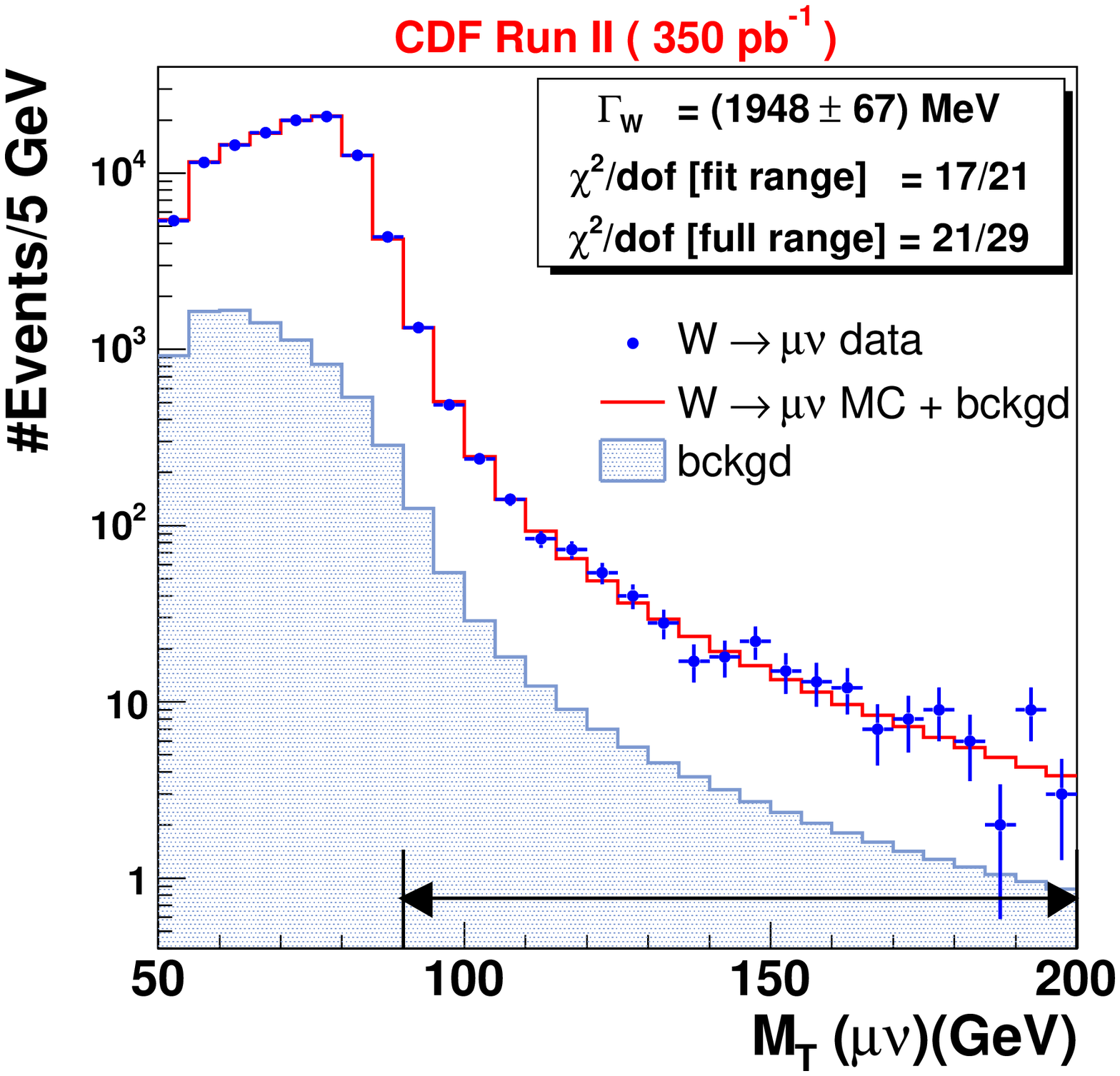}
\includegraphics[width=.495\textwidth]{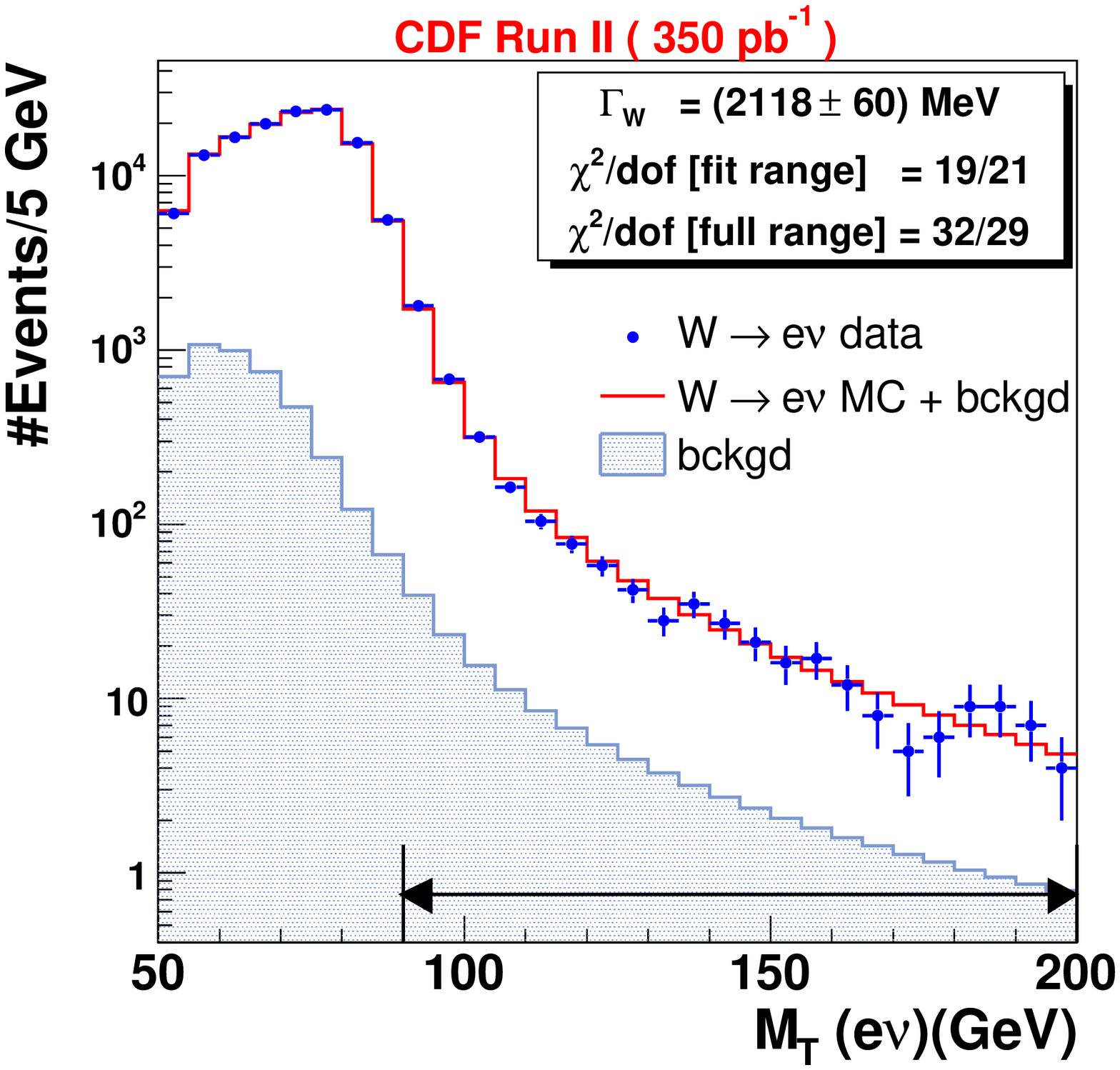}
\includegraphics[width=.495\textwidth]{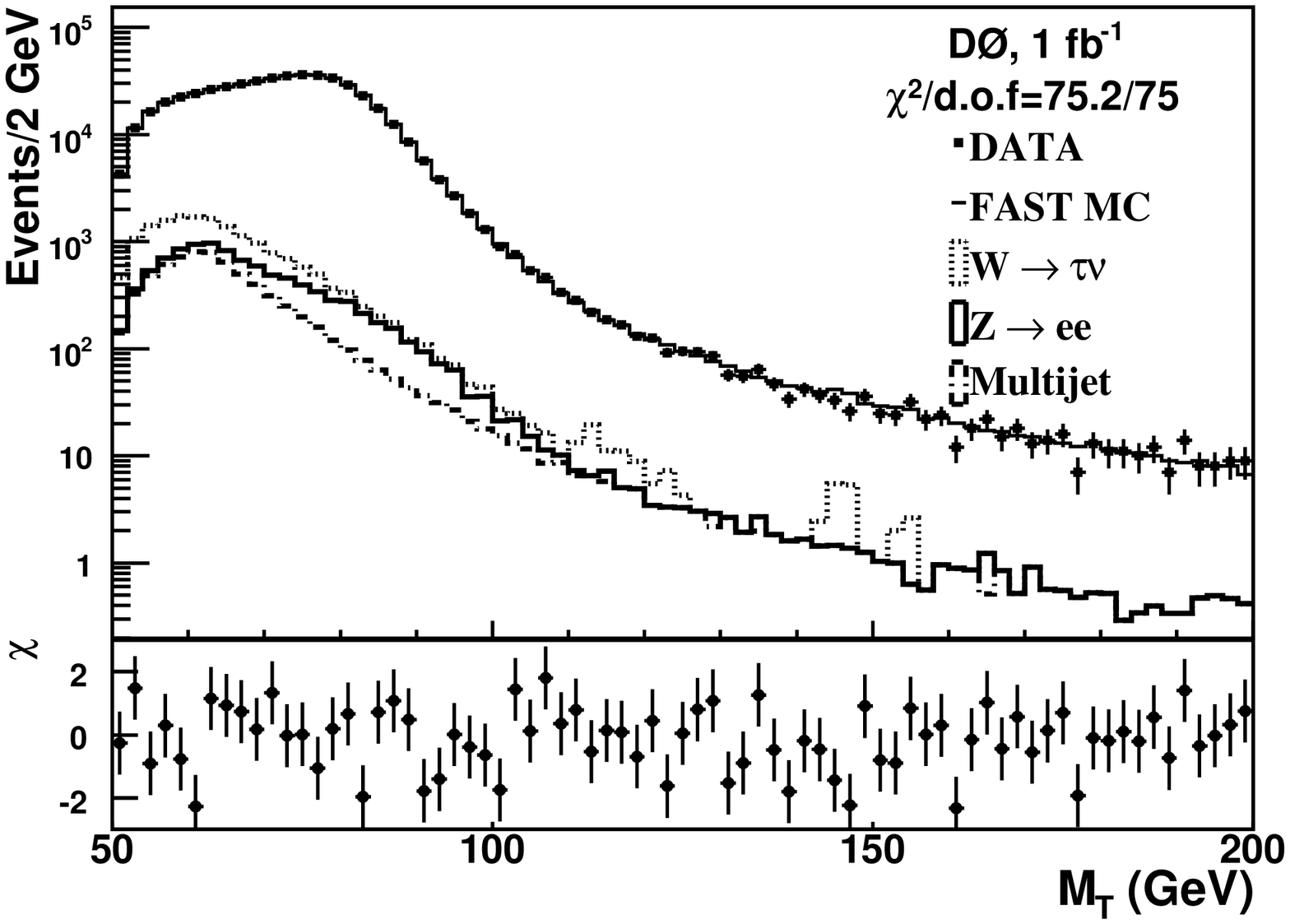}
\includegraphics[width=.495\textwidth]{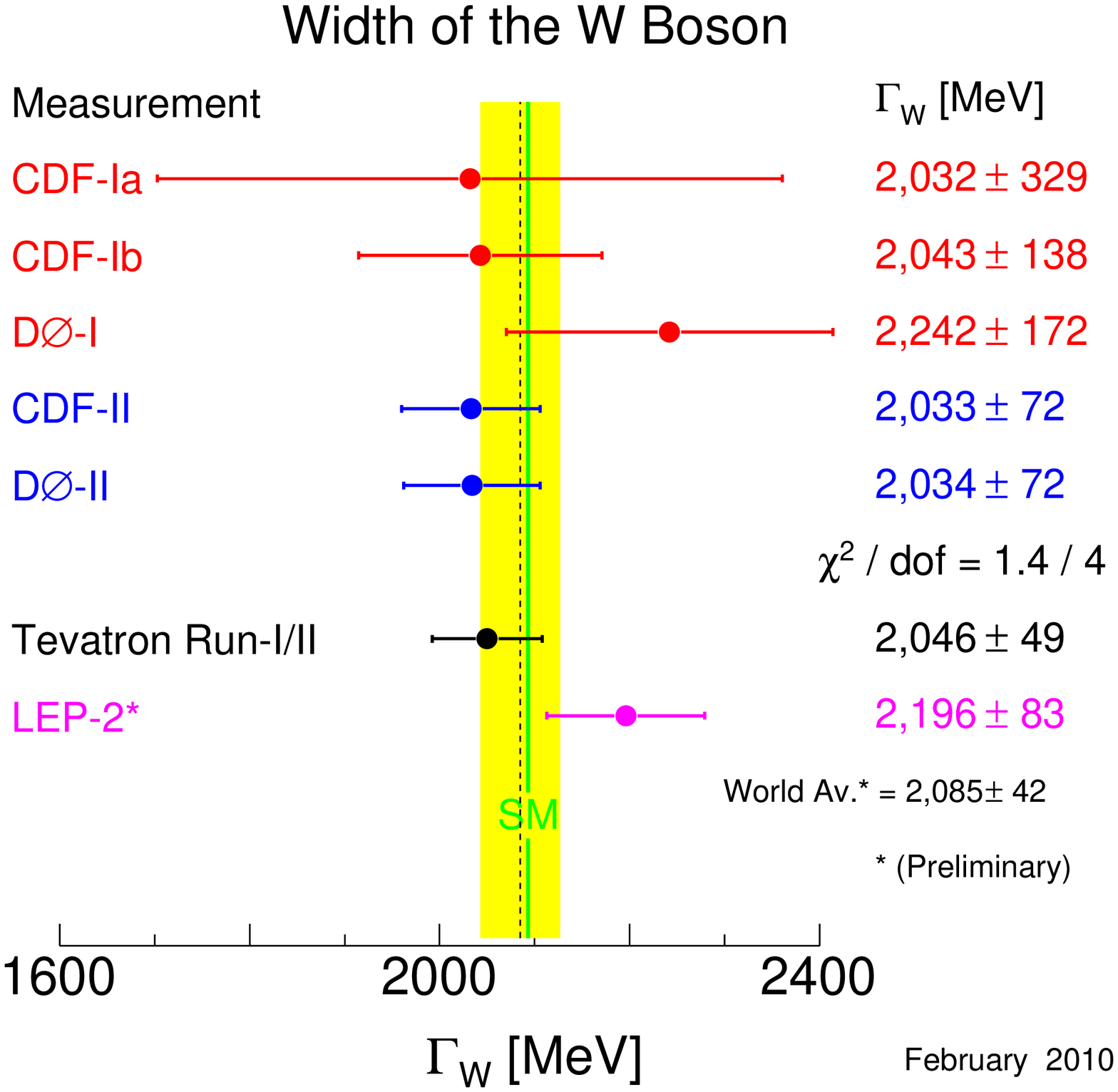}
\end{center}
\caption{Top left, top right, and bottom left: $M_T$ distributions for data 
and fast MC simulation with background added.
Two top plots: CDF. Bottom left plot from D0 shows also
signed $\chi$ values for each bin (bottom
part of the plot). Signed  $\chi$ is defined in the caption of Fig. 1. 
D0 used fitted $\Gamma_W$ value for the fast MC prediction rather
than the PDG value. The distribution of 
the fast MC simulation with background added is normalized to the number of data 
events in the region $50<M_T<100$ GeV (D0) and $50<M_T<90$ GeV (CDF).}
\label{fig:d0andcdfwidth}
\end{figure}
\begin{table}
  \caption{Uncertainties of D0 $\Gamma_W$ Measurement (MeV).}
  \label{tab:wgtable}
  \begin{tabular}{lc}
    \hline
   Source & $\Delta \Gamma_W$ (MeV) \\
  \hline
  Electron energy scale             & 33 \\
  Electron resolution model         & 10 \\
  Recoil model                      &  41  \\
  Electron efficiencies             &  19  \\
  Backgrounds                       &  6 \\ 
  PDF                               &  20 \\
  Electroweak radiative corrections &  7 \\
  Boson $p_T$                       &  1 \\ 
  $M_W$                             &  5 \\ \hline
  Total Systematic                  &  61 \\ \hline  
  Statistical                       &  39 \\ \hline
  Total                             &  72 \\ 
  \end{tabular}
\end{table}
\begin{table}
  \caption{Uncertainties of CDF $\Gamma_W$ Measurement (MeV).}
  \label{tab:cdfwgtable}
  \begin{tabular}{lccl}
  \hline
  Source                 & $e$ & $\mu$ &  common\\
  \hline 
  Lepton Scale     	 & 21 &  17 & 12 \\
  Lepton Resolution	 & 31 &  26 &  0 \\
  Simulation             & 13 &   0 &  0 \\
  Recoil                 & 54 &  49 &  0 \\
  Lepton ID              & 10 &   7 &  0 \\
  Backgrounds            & 32 &  33 &  0 \\
  $p_T(W)$               &  7 &   7 &  7 \\
  PDF                    & 20 &  20 & 20 \\
  QED    		 & 10 &   6 &  6 \\
  $M_W$                  &  9 &   9 &  9 \\ \hline
  Total Systematic       & 79 &   71 &  27 \\ \hline  
  Statistical            & 60 &   67 &  0 \\ \hline
  Total                  & 99 &   98 &  27 \\ 
  \end{tabular}
\end{table}
\newline
\newline
\newline
\newline
\newline

\end{document}